\documentstyle[11pt,newpasp,twoside]{article}
\def\edcomment#1{\iffalse\marginpar{\raggedright\sl#1\/}\else\relax\fi}
\reversemarginpar

\begin{document}
\title{$\omega$~Centauri as a Disrupted Dwarf Galaxy: Evidence from Multiple Stellar Populations}
\author{Young-Wook Lee, Soo-Chang Rey, Chang H. Ree, Jong-Myung Joo, Young-Jong Sohn, and Suk-Jin Yoon}
\affil{Center for Space Astrophysics, Yonsei University, Seoul 120-749, Korea}
\author{Alistair Walker}
\affil{National Optical Astronomy Observatories/Cerro Tololo Interamerican Observatory (NOAO/CTIO), Casilla 603, La Serena, Chile}

\begin{abstract}
Our recent CCD photometry (Lee et al. 1999) has shown, for the first time, that 
$\omega$~Cen has several  distinct stellar populations, which  is reminiscent of the 
Sagittarius dwarf galaxy.  Here we  present more  detailed analysis  of the  data 
along with the population  models. We confirm  the presence of  several distinct 
red-giant-branches (RGBs)  with a red metal-rich  sequence well separated  from 
other bluer   metal-poor ones.  Our population   models suggest  the red   clump 
associated with  the most  metal-rich  RGB is  about 4  Gyr  younger than  the 
dominant metal-poor component,  indicating that  $\omega$~Cen was enriched  over 
this timescale.  These features,  taken together  with this  cluster's other  unusual 
characteristics, provide good evidence that $\omega$~Cen was once  part of a more 
massive system that merged with the Milky Way, as the Sagittarius dwarf galaxy 
is in the process of doing  now. Mergers probably were much more  frequent in 
the early history of the  Galaxy and $\omega$~Cen  appears to be a  relict of this 
era. 
\end{abstract}

\section{Introduction}
As part of our investigation of the luminosity-metallicity relation of the RR Lyrae
stars in $\omega$~Cen, we have obtained $BV$ and $Ca$ \& Str\"omgren 
CCD frames with the CTIO 0.9m telescope that cover 40 $\times$ 40 $arcmin^2$ in a 3 $\times$ 3
grid centered on the cluster. As a byproduct of this investigation, we obtained 
high-quality homogeneous $BV$ color-magnitude (CM) data for more than 130,000 stars
in the field toward $\omega$~Cen, which represents the most extensive photometric survey
to date for this cluster as far as the stars brighter than the main-sequence (MS) turnoff
are concerned. This study (Lee et al. 1999) has shown, for the  first time, that
$\omega$~Cen has several distinct stellar populations, which is reminiscent of the
Sagittarius dwarf galaxy (Layden \& Sarajedini 1997). This feature was later confirmed
by Pancino et al. (2000) from  their independent photometry. In this paper, we present
our progress in more detailed analysis of the data along with the population models.

\section{Multiple Stellar Populations in $\omega$~Centauri}

\begin{figure*}
\includegraphics{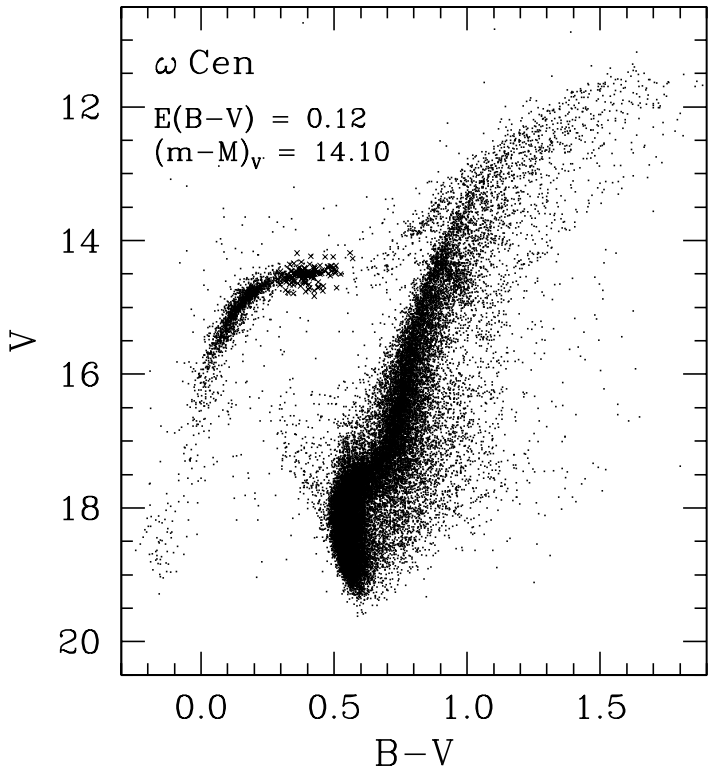}
\vspace{10.5cm}
\caption{Color-magnitude diagram of $\omega$~Cen from our $BV$ photometry.
The RR Lyrae variables are represented by crosses, and the foreground Galactic
disk star contamination is removed from our $Ca$ \& Str\"omgren $by$ photometry.}
\end{figure*}

Figure 1 presents $BV$ CM diagram for all stars in our program field including 
RR Lyrae variable stars. We were able to remove some foreground Galactic disk 
star contamination, as they are well discriminated from more metal-poor member 
stars of $\omega$~Cen in the analysis based on our $Ca$ \& Str\"omgren $by$ 
photometry (see Rey et al., this  volume). Here we confirm  our previous 
discovery that the red-giant-branch (RGB) of $\omega$~Cen has four distinct populations:
the most metal-poor sequence, metal-poor sequence, metal-rich sequence, and the most 
metal-rich sequence well separated from others. We believe this is 
a clear evidence for the multiple stellar populations in $\omega$~Cen.

The horizontal-branch (HB) distribution is also consistent with the discrete 
nature of RGB. We have the blue HB and RR Lyrae variables mainly associated 
with the two metal-poor RGBs, some red HB stars associated with the 
metal-rich population, and finally the red clump superimposed on the RGB, 
which must be associated with the most metal-rich population.

\section{Population Models}

\begin{figure*}
\includegraphics{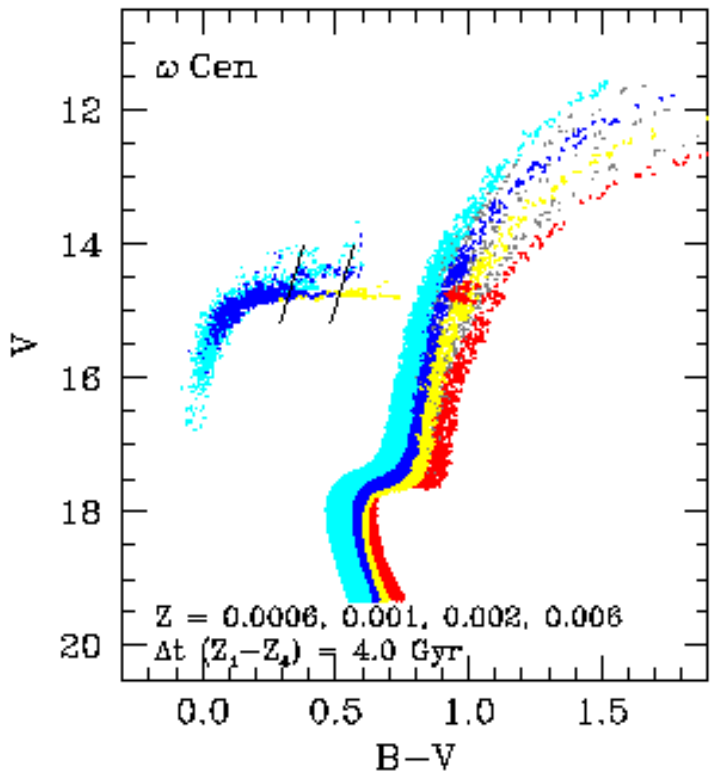}
\vspace{10.5cm}
\caption{Population model for $\omega$~Cen. Four distinct populations with different
metallicities can reproduce the observed discrete nature of the RGB and similar features
on the HB. The instability strip is schematically represented by solid lines.}
\end{figure*}

The presence of multiple stellar populations in $\omega$~Cen is confirmed 
by our population models  (Figure 2;  see also  Ree et  al., this  volume). Our 
models are constructed based on the  $Y^2$ isochrones (Yi et al. 2001) 
and corresponding HB evolutionary tracks (Yi, Demarque, \& Kim 1997) calculated with 
updated input physics. The reader is referred to Lee, Demarque, \& Zinn (1994) and
Park \& Lee (1997) for the details  of model constructions. From the population models, 
we confirm four distinct populations with different metallicities can reproduce the 
observed discrete  nature of the RGB. The models also confirm that the RR Lyrae 
variables are mainly produced by two metal-poor populations, with some contribution
from the 2nd most metal-rich population. 

For the RR Lyrae stars in $\omega$~Cen, we have new metallicity measurements from 
our $Ca$ \& Str\"omgren $by$ photometry (see Rey  et al., this volume), and therefore
more detailed comparison with the models is possible. 
From the luminosity-metallicity relation and the period-shift -- metallicity 
correlation of the RR Lyrae variables in $\omega$~Cen, we confrim that 
our models reproduce even the fine details within the instability strip, including 
the sudden upturn of the RR Lyrae luminosity and corresponding increase in 
period-shift at [Fe/H] $\sim$ -1.5, which is a result of redward evolution from the 
blue HB (see Rey et al., this volume; Yoon \& Lee, this volume). This suggests 
we have a very good understanding of what is happening in the instability strip 
of $\omega$~Cen.

\begin{figure*}
\includegraphics{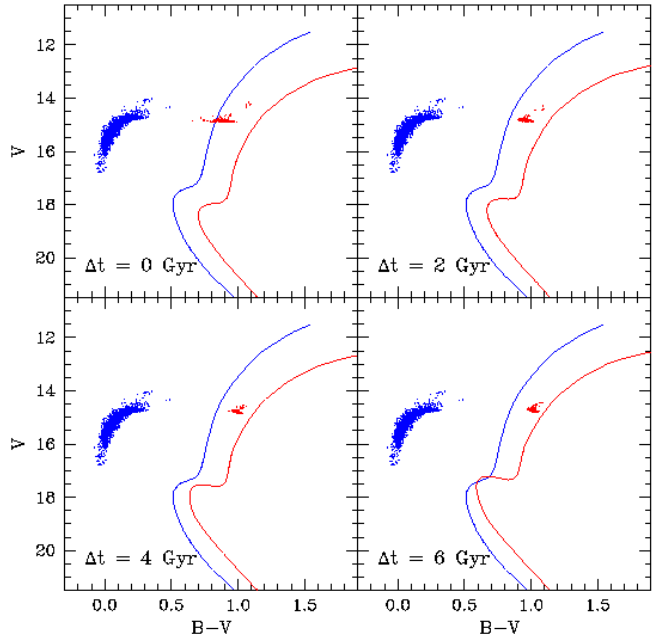}
\vspace{9cm}
\caption{The models illustrate the estimation of age difference between the red clump
associated with the most metal-rich population and the blue HB associated with the
most metal-poor component.}
\end{figure*}

Our models also reproduce the red clump associated with the most metal-rich 
RGB, but only when the age  of the most metal-rich population is some 4 Gyr 
younger than the most metal-poor population. This is illustrated in Figure  3, 
where we can see the variation of the  red clump location in  the CM diagram 
under different assumptions regarding the age difference betwen the most 
metal-rich and the most metal-poor populations. By comparing these models with 
the observed color of the red clump, which is estimated from the analysis of the 
luminosity functions of the RGBs superimposed on the red clump (see Rey  et al., 
this volume), we conclude that the $\Delta$t of $\sim$ 4 Gyrs is our best 
estimate for the age spread within the $\omega$~Cen, in the sense that the 
metal-rich population is younger. This is in qualitative agreement with the results 
obtained directly from the Str\"omgren photometry of main-sequence stars
(Hughes \& Wallerstein 2000; Hilker \& Richtler 2000), although these results
are subject to large errors associated with the photometry and metallicity groupings.

These unusual characteristics of $\omega$~Cen are very similar to the 
case of Sagittarius dwarf galaxy, which is in the process of disruption with  the 
Milky Way (Ibata, Gilmore, \& Irwin 1994). In Figure 4, we have compared our population 
models with the observations centered on the M54 (Layden \& Sarajedini 2000), which is 
believed to be the nucleus of the Sagittarius dwarf galaxy. Note again that the 
blue HB is from the most metal-poor (and older) population, the red HB is 
from the intermediate metallicity population, and finally the red clump is from 
the most metal-rich (and younger) population. We confrim here that the 
Sagittarius dwarf galaxy also has several distinct stellar populations with the 
internal age-metallicity relation that spans $\sim$ 7 Gyrs. Other Local Group dwarf 
galaxies are also known to have similarly complex star formation histories (Mateo 1998). 

\begin{figure*}
\includegraphics{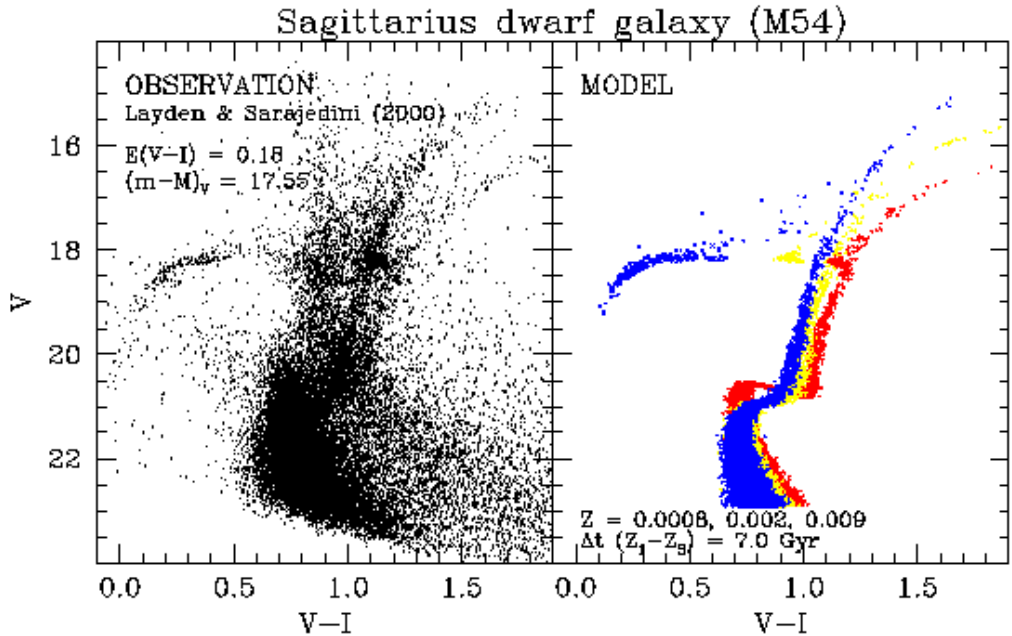}
\vspace{9cm}
\caption{Comparison of observation and population model for the Sagittarius dwarf galaxy.}
\end{figure*}

\section{Discussion}

\subsection{The Origin of $\omega$~Centauri}
The multiple populations and the internal age-metallicity relation found in our 
work (and also suggested by others in this workshop), which resemble the 
characteristics of Sagittarius dwarf galaxy, suggest $\omega$~Cen was 
massive enough for self-enrichment and several bursts of star formation. The 
relatively extended enrichment period of $\sim$ 4 Gyrs then indicates that the initial 
evolution of $\omega$~Cen occured away from the dense central regions of 
young Milky Way, because, near the Galactic center like the current location of 
$\omega$~Cen, we would expect the gaseous materials to have been 
stripped from the proto $\omega$~Cen on a much shorter timescale as a 
result of disk shocking and/or similar processes. This view is not inconsistent 
with $\omega$~Cen's rather unusual kinematics and orbit (Dinescu, this 
volume). 

From the discrete nature of RGB, some may suggest that the $\omega$~Cen
could be a merger of several globular clusters. But mergers of (at least) 
four clusters are very unlikely, if not impossible, in the Milky Way, considering 
the rather high velocities between the clusters in the halo. We can not rule out 
this scenario in dwarf galaxies, however, because internal velocity dispersion is 
much lower in dwarf galaxies (van den Bergh 1996). 

All of these information strongly suggest that the $\omega$~Cen is a 
relict or a nucleus of a disrupted dwarf galaxy. The case of $\omega$~Cen 
and that of the Sagittarius dwarf system therefore provide direct evidence for 
past and continuing accretion of protogalactic fragments, which suggest that 
similar accretion events may have continued throughout the history of Milky 
Way formation.

\subsection{Other Globular Clusters with Multiple Populations?}
It is interesting to note that two globular clusters, now identified as the nuclei 
or parts of disrupted dwarf galaxies, are among the most massive globular 
clusters in the Milky Way. $\omega$~Cen is, of course, the most massive 
globular cluster, and M54, the nucleus of the Sagittarius dwarf, is the 2nd most 
massive globular cluster. Then, how about the other massive globular clusters? 
There are several Galactic globular clusters that resemble $\omega$~Cen, 
but among them, we found the cases of NGC~6388 and NGC~6441 are the 
most interesting. They are 3rd and 5th most massive globular clusters, 
respectively, and they all have very peculair bimodal HB distributions. Our 
population models for these unusual clusters show that the adoption of two 
distinct populations within the systems and very mild internal age-metallicity 
relations between the two populations can reproduce the observed features on the 
HB and RGB (see Ree et al., this volume). This conclusion is also supported by 
the fact that the mean period of RR Lyrae variables in these clusters are too 
long for their high metallicities, which is understood in two populations secnario, 
where the RR Lyraes are highly evolved stars from the older and metal-poor 
blue HB population (see Ree et al., this volume). 

This suggests some massive globular clusters in the Milky Way are probably 
not genuine globular clusters, but infact relicts of disrupted dwarf galaxies like 
$\omega$~Cen and M54.  

\subsection{Classification of Galactic Globular Cluster System}
Our conclusion on the origin of $\omega$~Cen and other massive globular 
clusters, when combined with other recent findings on the origin of Galactic 
globular cluster system, suggest that the present day Galactic globular clusters 
are infact subdivided into three different types:

(a) Clusters formed in a collapsing proto-Galaxy (e.g., Eggen, Lyndel-Bell, \& Sandage 1962).
They are mostly in the inner halo and are only genuine Galactic globular clusters.
 
(b) Clusters originally formed in satellite dwarf galaxies later accreted to the 
Milky Way. They include globular clusters (Pal~12, Ter~7, Ter~8, \& Arp~2) belong 
to Sagittarius dwarf galaxy, and young outer halo clusters with retrograde motion 
(Zinn 1993; van den Bergh 1993). Yoon \& Lee (this volume) also reported very 
compelling evidence that metal-poor ([Fe/H] $\la$ -2.0) Oosterhoff II clusters have 
the positional and orbital characteristics fully consistent with the hypothesis that 
they were originated from a satellite galaxy.

(c) Nuclei (or relicts) of disrupted dwarf galaxies. $\omega$~Cen and M54, 
the nucleus of the Sagittarius dwarf, are undoubtedly belong to this type. Our 
population models (Ree et al., this volume) suggest other massive globular 
clusters with bimodal HBs (e.g., NGC~6388 and NGC~6441) may also belong to 
this category.

\end{document}